\documentclass[preprint,prd,nofootinbib,tightenlines,amsmath,amssymbol]{revtex4}
\usepackage{graphicx}
\usepackage{bm}

\begin{document}
\baselineskip=15pt \parskip=5pt

\vspace*{3em}

\preprint{}

\title{Penguin and Box Diagrams in Unitary Gauge}

\author{Xiao-Gang He}
\email{hexg@phys.ntu.edu.tw}
\affiliation{Department of Physics and Center for Theoretical Sciences, \\
National Taiwan University, Taipei 106, Taiwan}

\author{Jusak Tandean}
\email{jtandean@yahoo.com}
\affiliation{Department of Physics and Center for Theoretical Sciences, \\
National Taiwan University, Taipei 106, Taiwan}

\author{G. Valencia}
\email{valencia@iastate.edu}
\affiliation{Department of Physics and Astronomy, Iowa State University, Ames, Iowa 50011, USA}

\date{\today $\vphantom{\bigg|_{\bigg|}^|}$}

\begin{abstract}

We evaluate one-loop diagrams in the unitary gauge that contribute to flavor-changing neutral
current (FCNC) transitions involving two and four fermions.
Specifically, we deal with penguin and box diagrams arising within the standard model (SM) and
in nonrenormalizable extensions thereof with anomalous couplings of the $W$ boson to quarks.
We show explicitly in the SM the subtle cancelation among divergences from individual
unitary-gauge contributions to some of the physical FCNC amplitudes and
derive expressions consistent with those obtained using $R_\xi$ gauges in the literature.
Some of our results can be used more generally in certain models involving fermions and gauge
bosons which have interactions similar in form to those we consider.

\end{abstract}

\pacs{PACS numbers: }

\maketitle

In the presence of new physics affecting primarily the charged weak currents involving quarks,
the $W$ boson may have couplings to quarks beyond those in the standard model (SM).
The effective Lagrangian for a~general parametrization of the $W$ boson interacting with
an up-type quark $U$ and a~down-type quark $D$ can be written as
\begin{eqnarray} \label{Ludw}
{\cal L}_{UDW}^{} \,\,=\,\,  -\bar U\gamma^\mu \bigl(
g_{\rm L}^D P_{\rm L}^{} \,+\, g_{\rm R}^D P_{\rm R}^{} \bigr) D\, W_\mu^+
\,\,+\,\, {\rm H.c.} ~,
\end{eqnarray}
where $g_{\rm L,R}^D$ are complex coupling constants and
\,$P_{\rm L,R}^{}=\frac{1}{2}(1\mp\gamma_5^{})$.\,
In the SM limit, these constants become
\begin{eqnarray}
g_{\rm L}^D \,\,=\,\, \frac{g}{\sqrt2}\, V_{UD}^{} ~, \hspace{7ex}
g_{\rm R}^D \,\,=\,\, 0 ~,
\end{eqnarray}
where $g$ is the weak coupling constant and $V_{kl}$ are elements of
the Cabibbo-Kobayashi-Maskawa (CKM) matrix.

Beyond the SM, the new couplings in Eq.~(\ref{Ludw}) not only affect weak decays through
tree-level interactions, but also modify flavor-changing neutral current (FCNC) transitions
at one-loop level.
In Ref.~\cite{He:2009hz}, we have studied several two- and four-fermion FCNC transitions
induced at one-loop level by the new couplings in the charm sector and determined constraints
on them from various processes.
Here we provide the detailed derivation of the loop formulas summarized therein.
Some of the results we present in this paper can also be applied to other models involving
fermions and gauge bosons which have interactions similar in form to those we consider.

In treating the loops, we adopt the unitary gauge, which is convenient in the absence of
knowledge about the new degrees of freedom.
Another advantage is that, since diagrams involving unphysical states are absent in unitary
gauge, the number of diagrams to deal with is smaller than that in the $R_\xi$ gauges usually
employed in the literature.
Unlike in $R_\xi$ gauges, our calculations in unitary gauge produce divergences in
some of the individual contributions.
To treat them, we employ dimensional regularization with a fully anticommuting~$\gamma_5^{}$.
We express the divergent part of the $D$-dimensional integration in terms of the combination
\begin{eqnarray} \label{delta}
\Delta \,\,=\,\, \frac{2}{4-D} - \gamma_{\rm E}^{} \,+\, \ln\frac{4\pi\, \mu^2}{m_W^2} ~,
\end{eqnarray}
where  $\gamma_{\rm E}^{}$  is the Euler constant and $\mu$ the mass scale that arises from
the regularization procedure.
For physical processes, the divergent parts should cancel in the total amplitudes in the SM
limit and the results should agree with those obtained in $R_\xi$ gauges.
It is therefore instructive to see how the individual contributions in unitary gauge differ
from those in the general $R_\xi$ gauge.

\begin{figure}[t]
\includegraphics{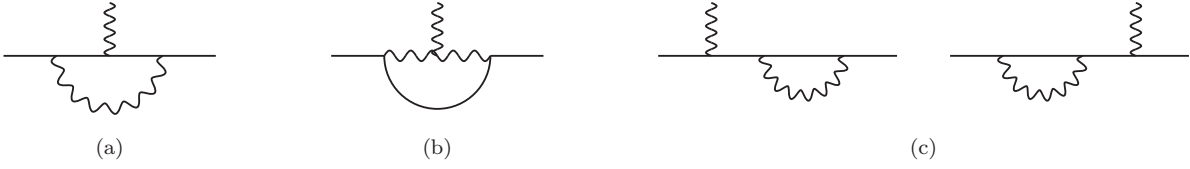}
\caption{\label{penguins}Diagrams contributing to amplitudes for \,$d\to d'{\cal V}^*$,\, with
$\cal V$ being a neutral gauge boson. In all figures, straight lines denote fermions and
the loops contain $W$ bosons besides fermions.}
\end{figure}

\begin{figure}[ht]
\includegraphics{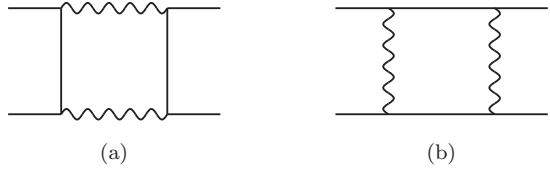}
\caption{\label{boxes}Box diagrams contributing to amplitudes for (a)~$d\bar d'\to\nu\bar\nu$ or
$\ell^+\ell^-$\, and (a,b)~$d\bar d'\to d\bar d'$.}
\end{figure}

At the quark level, the loop-induced physical processes of interest in Ref.~\cite{He:2009hz}
are  \,$d\to d'\gamma$,\, \,$d\to d'g$,\, \,$d\bar d'\to\nu\bar\nu$,\,
\,$d\bar d'\to\ell^+\ell^-$,\, and  $d\bar d'\to\bar d d'$,\,  where  $d$ and $d'$ are
down-type quarks, with $\gamma$, $g$, $\nu$, and $\ell$ as usual denoting a photon, gluon,
neutrino, and charged lepton, respectively.
The relevant diagrams are displayed in Figs.~\ref{penguins} and~\ref{boxes}, where
the loops in unitary gauge contain only fermions and $W$ bosons.
The \,$d\bar d'\to\nu\bar\nu$\, and \,$d\bar d'\to\ell^+\ell^-$\, transitions also receive
$Z$-penguin contributions and, for the latter, $\gamma$-penguin contributions as well.


Some of our results are new, while some have been found before using unitary or $R_\xi$ gauges.
Specific aspects of our paper not previously available in the literature include
unitary-gauge expressions for individual amplitudes corresponding to the separate diagrams
in Fig.~\ref{penguins} contributing to \,$d\to d'{\cal V}^*$,\, arising from the quark-$W$
interactions of Eq.~(\ref{Ludw}) and SM-like couplings of the neutral gauge bosons~$(Z,\gamma,g)$
to fermions and the~$W$.
These particular results, presented in the Appendix, can thus be used more generally in models
involving fermions and gauge bosons having interactions similar in form to those we discuss.
Another new aspect of our work is the explicit demonstration, in unitary gauge, of cancelation
between divergences from penguin and box diagrams contributing to
\,$d\bar d'\to\nu\bar\nu$\,  and  \,$d\bar d'\to\ell^+\ell^-$.\,
Moreover, perhaps more completely than earlier work in the literature, we provide a~set of
unitary-gauge results for the FCNC transitions listed above induced by the interactions in
Eq.~(\ref{Ludw}), along with SM-like interactions of the neutral gauge bosons.
We will compare our results with those in the literature, where available.

In the following, we derive first the amplitudes for \,$d\to d'{\cal V}^*$,\,  where
$d$ and $d'$ are on-shell, but the neutral gauge-boson ${\cal V}$ is off-shell with momentum~$k$
that is small compared to $m_W^{}$.
We have collected the resulting formulas corresponding to the diagrams in Fig.~\ref{penguins}(a),
(b), and~(c) in the Appendix.
Then, assuming that $\cal V$ has SM interactions with fermions, the couplings being given in
Eq.~(\ref{sm}), we combine the three contributions.
Thus we obtain for \,${\cal V}=Z$\,
\begin{eqnarray} \label{ddz}
{\cal M}_{d\to d'Z^*}^q \,\,=\,\,
\frac{g\,\varepsilon_\mu^*}{16\pi^2\, c_{\rm w}^{}}\, \bar d' \Biggl[
\gamma^\mu\bigl(z_{\rm L}^q P_{\rm L}^{} + z_{\rm R}^q P_{\rm R}^{}\bigr) +
\frac{i\sigma^{\mu\nu}k_\nu^{}}{m_W^2}\bigl(Z_{\rm L}^q P_{\rm L}^{}+Z_{\rm R}^q P_{\rm R}^{}\bigr)
\Biggr] d ~,
\end{eqnarray}
upon setting set \,$k^2=0$,\, where
\begin{eqnarray} \label{z}
\begin{array}{c}   \displaystyle
z_{\rm L}^q \,\,=\,\, \frac{g_{\rm L}^d\, \bar g_{\rm L}^{d'}}{4} \Biggl[
-x_q^{}\,\Delta \,+\, \frac{7 x_q^{}-x_q^2}{2\bigl(1-x_q^{}\bigr)} +
\frac{4 x_q^{}-2 x_q^2+x_q^3}{\bigl(1-x_q^{}\bigr)^2}\, \ln x_q^{} \Biggr] ~,
\vspace{2ex} \\   \displaystyle
z_{\rm R}^q \,\,=\,\, \frac{g_{\rm R}^d\, \bar g_{\rm R}^{d'}}{4} \Biggl[
7 x_q^{}\,\Delta \,+\, \frac{15 x_q^{}-9 x_q^2}{2\bigl(1-x_q^{}\bigr)} -
\frac{4 x_q^{}-14 x_q^2+7 x_q^3}{\bigl(1-x_q^{}\bigr)^2}\, \ln x_q^{} \Biggr] ~,
\end{array}
\end{eqnarray}
\begin{eqnarray}
\begin{array}{c}   \displaystyle
Z_{\rm L}^q \,\,=\,\,
g_{\rm L}^d\, \bar g_{\rm L}^{d'}\, m_{d'}^{}\, {\cal Z}_0^{\rm SM}\big(x_q^{}\bigr) +
g_{\rm R}^d\, \bar g_{\rm R}^{d'}\, m_d^{}\, z_0^{}\big(x_q^{}\bigr) \,+\,
g_{\rm L}^d\,\bar g_{\rm R}^{d'}\, m_q^{}\, {\cal Z}_0^{}\big(x_q^{}\bigr) ~,
\vspace{2ex} \\   \displaystyle
Z_{\rm R}^q \,\,=\,\,
g_{\rm L}^d\, \bar g_{\rm L}^{d'}\, m_d^{}\, {\cal Z}_0^{\rm SM}\big(x_q^{}\bigr) +
g_{\rm R}^d\, \bar g_{\rm R}^{d'}\, m_{d'}^{}\, z_0^{}\big(x_q^{}\bigr) \,+\,
\bar g_{\rm L}^{d'} g_{\rm R}^d\, m_q^{}\, {\cal Z}_0^{}\big(x_q^{}\bigr) ~,
\end{array}
\end{eqnarray}
\begin{eqnarray} \label{z0}
\begin{array}{c}   \displaystyle
{\cal Z}_0^{\rm SM}(x) \,\,=\,\, \frac{13 x-2 x^2+x^3}{24(1-x)^3}
- \frac{7 x-5 x^2-8 x^3}{12(1-x)^3}\,c_{\rm w}^2 \,+\,
\frac{x+x^2-\bigl(4 x^2-6 x^3\bigr)c_{\rm w}^2}{4(1-x)^4}\, \ln x ~,
\vspace{2ex} \\   \displaystyle
z_0^{}(x) \,\,=\,\, {\cal Z}_0^{\rm SM}(x) \,-\,
\frac{16 x-29 x^2+7 x^3}{24(1-x)^3} - \frac{2 x-3 x^2}{4(1-x)^4}\, \ln x ~,
\vspace{2ex} \\   \displaystyle
{\cal Z}_0^{}(x) \,\,=\,\,
\frac{5\bigl(4+x+x^2\bigr)-4\bigl(20-31 x+5 x^2\bigr)c_{\rm w}^2}{24(1-x)^2} \,+\,
\frac{5 x-4\bigl(2 x-3 x^2\bigr)c_{\rm w}^2}{4(1-x)^3}\, \ln x ~.
\end{array}
\end{eqnarray}
Evidently, the current $(\gamma^\mu)$ terms are divergent, whereas the magnetic
$(\sigma^{\mu\nu})$ terms are finite.  Within the SM, the term containing $z_{\rm L}^q$ has been
calculated previously using unitary gauge in Refs.~\cite{Buras:2006wk,Wu:2006sp}, and our
expression above agrees with the results found therein.

In writing down ${\cal M}_{d\to d'Z^*}^q$, we have dropped finite terms (without $\Delta$) that
do not depend on the internal quark mass $m_q^{}$, or equivalently \,$x_q^{}=m_q^2/m_W^2$.\,
This is because such terms will be removed by the Glashow-Iliopoulos-Maiani (GIM) mechanism
in the SM, after summing over \,$q=u,c,t$\,  and imposing the unitarity relation
\,$V_{ud'}^*V_{ud}^{}+V_{cd'}^*V_{cd}^{}+V_{td'}^*V_{td}^{}=0$,\,  and by a GIM-like mechanism
beyond the SM.  For the same reason, we also drop $m_q^{}$-independent finite terms in the next
two amplitudes for \,$d\to d'\cal V^*$,\, as well as in the box-diagram contributions below.

For \,${\cal V}=\gamma$,\, after combining the three contributions (a,b,c), we find that
the $k$-independent terms cancel completely, but that terms of first and second orders in~$k$
do not, which we keep, namely
\begin{eqnarray} \label{ddf}
{\cal M}_{d\to d'\gamma^*}^q \,\,=\,\,
\frac{e\, \varepsilon_\mu^*}{16\pi^2\, m_W^2}\, \bar d' \Bigl[
\bigl(k^2 \gamma^\mu-\!\!\not{\!k} k^\mu\bigr)
\bigl(f_{\rm L}^q P_{\rm L}^{} + f_{\rm R}^q P_{\rm R}^{}\bigr)
+ i\sigma^{\mu\nu}k_\nu^{} \bigl(F_{\rm L}^q P_{\rm L}^{}+F_{\rm R}^q P_{\rm R}^{}\bigr)
\Bigr] d ~,
\end{eqnarray}
where
\begin{eqnarray}
f_{\rm L}^q &=&
\frac{g_{\rm L}^d\, \bar g_{\rm L}^{d'}}{4} \Biggl[ \biggl(x_q^{}-\frac{16}{3}\biggr) \Delta
\,+\, \frac{153 x_q^{}-383 x_q^2+245 x_q^3-27 x_q^4}{18\bigl(1-x_q^{}\bigr)^3}
\nonumber \\ && \hspace*{7ex} +\,\,
\frac{-16+64 x_q^{}-36 x_q^2-93 x_q^3+84 x_q^4-9 x_q^5}{9\bigl(1-x_q^{}\bigr)^4}\, \ln x_q^{}
\Biggr] ~, \\
f_{\rm R}^q &=&
\frac{g_{\rm R}^d\, \bar g_{\rm R}^{d'}}{g_{\rm L}^d\, \bar g_{\rm L}^{d'}}\,f_{\rm L}^q ~,
\end{eqnarray}
\begin{eqnarray} \label{FLR}
\begin{array}{c}   \displaystyle
F_{\rm L}^q \,\,=\,\,
2\bigl(g_{\rm L}^d\,\bar g_{\rm L}^{d'}\,m_{d'}^{}+g_{\rm R}^d\,\bar g_{\rm R}^{d'}\,m_d^{}\bigr)\,
F_0^{\rm SM}\big(x_q^{}\bigr) \,+\,
2\,g_{\rm L}^d\,\bar g_{\rm R}^{d'}\, m_q^{}\, F_0^{}\big(x_q^{}\bigr) ~,
\vspace{2ex} \\   \displaystyle
F_{\rm R}^q \,\,=\,\,
2\bigl(g_{\rm L}^d\,\bar g_{\rm L}^{d'}\,m_d^{}+g_{\rm R}^d\,\bar g_{\rm R}^{d'}\,m_{d'}^{}\bigr)\,
F_0^{\rm SM}\big(x_q^{}\bigr) \,+\,
2\,\bar g_{\rm L}^{d'} g_{\rm R}^d\, m_q^{}\, F_0^{}\big(x_q^{}\bigr) ~,
\end{array}
\end{eqnarray}
\begin{eqnarray} \label{f}
\begin{array}{c}   \displaystyle
F_0^{\rm SM}(x) \,\,=\,\,
\frac{-7 x+5 x^2+8 x^3}{24(1-x)^3} \,-\, \frac{2 x^2-3 x^3}{4(1-x)^4}\,\ln x ~,
\vspace{2ex} \\   \displaystyle
F_0^{}(x) \,\,=\,\, \frac{-20+31 x-5 x^2}{12(1-x)^2} \,-\, \frac{2x-3x^2}{2(1-x)^3}\, \ln x ~.
\end{array}
\end{eqnarray}
This amplitude also contains terms which are divergent, but its magnetic part is finite, as in
the \,${\cal V}=Z$\, case.  The $f_{\rm L}^q$ term has been evaluated in unitary gauge
before~\cite{Buras:2006wk,Wu:2006sp,Chia:1985dx}, and our result confirms those given in
Refs.~\cite{Buras:2006wk,Wu:2006sp},\footnote{A factor of $x_t^{}$ seems to be missing from
the $T^{\rm div}$ term in the Eq.~(25) of Ref.~\cite{Wu:2006sp}} but differs from that in
Ref.~\cite{Chia:1985dx}.
The magnetic, $F_{\rm L,R}^q$, terms have also been evaluated in unitary gauge previously in
Refs.~\cite{Chia:1985dx,Fujikawa:1993zu}, and the resulting expressions are consistent with
ours, at linear order in~$g_{\rm R}^{}$.

For $\cal V$ being a gluon, $g_a^{}$, the diagram in Fig.~\ref{penguins}(b) is absent, and we get
\begin{eqnarray} \label{ddg}
{\cal M}_{d\to d'g_a^*}^q \,\,=\,\,
\frac{g_{\rm s}^{}\, \varepsilon_\mu^*}{16\pi^2\, m_W^2}\, \bar d' \Bigl[
\bigl(k^2 \gamma^\mu-\!\!\not{\!k} k^\mu\bigr)
\bigl(g_{\rm L}^q P_{\rm L}^{} + g_{\rm R}^q P_{\rm R}^{}\bigr)
+ i\sigma^{\mu\nu}k_\nu^{} \bigl(G_{\rm L}^q P_{\rm L}^{}+G_{\rm R}^q P_{\rm R}^{}\bigr) \Bigr]
t_a^{} d ~,
\end{eqnarray}
where
\begin{eqnarray}
g_{\rm L}^q \,\,=\,\,
\frac{g_{\rm L}^d\, \bar g_{\rm L}^{d'}}{g_{\rm R}^d\, \bar g_{\rm R}^{d'}}\,g_{\rm R}^q
\,\,=\,\, \frac{g_{\rm L}^d\, \bar g_{\rm L}^{d'}}{4} \Biggl[
\frac{18 x_q^{}-11 x_q^2-x_q^3}{3\bigl(1-x_q^{}\bigr)^3} \,-\,
\frac{8-32 x_q^{}+18 x_q^2}{3\bigl(1-x_q^{}\bigr)^4}\,\ln x_q^{} \Biggr] ~,
\end{eqnarray}
\begin{eqnarray} \label{GLR}
\begin{array}{c}   \displaystyle
G_{\rm L}^q \,\,=\,\,
2\bigl(g_{\rm L}^d\,\bar g_{\rm L}^{d'}\,m_{d'}^{}+g_{\rm R}^d\,\bar g_{\rm R}^{d'}\,m_d^{}\bigr)\,
G_0^{\rm SM}\big(x_q^{}\bigr) \,+\,
2\,g_{\rm L}^d\,\bar g_{\rm R}^{d'}\, m_q^{}\, G_0^{}\big(x_q^{}\bigr) ~,
\vspace{2ex} \\   \displaystyle
G_{\rm R}^q \,\,=\,\,
2\bigl(g_{\rm L}^d\,\bar g_{\rm L}^{d'}\,m_d^{}+g_{\rm R}^d\,\bar g_{\rm R}^{d'}\,m_{d'}^{}\bigr)\,
G_0^{\rm SM}\big(x_q^{}\bigr) \,+\,
2\,\bar g_{\rm L}^{d'} g_{\rm R}^d\, m_q^{}\, G_0^{}\big(x_q^{}\bigr) ~,
\end{array}
\end{eqnarray}
with
\begin{eqnarray}  \label{g}
G_0^{\rm SM}(x) \,\,=\,\,
\frac{-2 x-5 x^2+x^3}{8(1-x)^3} \,-\, \frac{3 x^2\,\ln x}{4(1-x)^4} ~,
\hspace{5ex}
G_0^{}(x) \,\,=\,\, \frac{-4-x-x^2}{4(1-x)^2} \,-\, \frac{3x\, \ln x}{2(1-x)^3} ~.
\end{eqnarray}
Unlike in the \,${\cal V}=Z$ and $\gamma$\, cases, this amplitude has no divergence.
Our expressions for the terms containing $g_{\rm L}^q$ and $G_0^{\rm SM}$ in the SM limit agree
with the corresponding unitary-gauge results of Ref.~\cite{Chia:1983hd}, except for the relative
sign between them.
The magnetic part containing $G_0^{}$ has been previously calculated in unitary gauge in
Ref.~\cite{AbdElHady:1997eu}, but the result therein misses a factor of~$\frac{1}{2}$.

Before proceeding to evaluate the box diagrams, we note that the divergent terms in
${\cal M}_{d\to d'Z^*}^q$ and ${\cal M}_{d\to d'\gamma^*}^q$ above depend on $x_q^{}$.
This implies that in the SM limit,
\begin{eqnarray} \label{smlimit}
g_{\rm L}^d \,\,=\,\, \frac{g}{\sqrt2}\, V_{qd}^{} ~, \hspace{5ex}
\bar g_{\rm L}^{d'} \,\,=\,\, \frac{g}{\sqrt2}\, V_{qd'}^* ~, \hspace{5ex}
g_{\rm R}^d \,\,=\,\, g_{\rm R}^{d'} \,\,=\,\, 0 ~,
\end{eqnarray}
these divergences cannot be eliminated by the GIM mechanism.
We have checked that such divergences would still be present in the absence of
the second diagram in Fig.~\ref{penguins}(b), which has a greater degree of divergence than
the others.
In contrast, the corresponding terms calculated in $R_\xi$ gauges are
finite~\cite{Ma:1979px,Inami:1980fz,Deshpande:1981zq}.
The \,$k^2\gamma^\mu-\!\!\not{\!k}k^\mu$\, parts of the \,${\cal V}=Z,\gamma$\, amplitudes
are, therefore, gauge dependent.
For physical processes, such as \,$d\bar d'\to\nu\bar\nu$\, and \,$d\bar d'\to\ell^+\ell^-$,\,
we will show later that the $x_q^{}$-dependent divergences contributed by
the \,$d\to d'{\cal V}^*$\, diagrams in unitary gauge are canceled exactly by
the $x_q^{}$-dependent divergences in the box-diagram contributions.
The magnetic $(\sigma^{\mu\nu})$ terms in the \,$d\to d'{\cal V}^*$\, amplitude are,
on the other hand, gauge independent.
In particular, our expressions for the $F_{\rm L,R}^q$ and $G_{\rm L,R}^q$ terms (at first
order in~$g_{\rm R}^{}$) agree with those derived using $R_\xi$
gauges~\cite{Inami:1980fz,Deshpande:1981zq,Cho:1993zb}

We also note that the divergent terms in ${\cal M}_{d\to d'Z^*}^q$ will also be present in
the amplitude for the physical decay  \,$Z\to\bar d d'$.\,  In that case, $k^2$-dependent
terms need to be kept and, as a consequence, there are additional divergent contributions.
The divergent part of the \,$d\to d'Z^*$\, amplitude in the \,$k^2\neq0$\, case is then
\begin{eqnarray} \label{Dddz}
{\cal M}_{d\to d'Z^*}^{q,\rm(div)} &=&
\frac{g\,\Delta}{16\pi^2}\, \bar d'\!\!\not{\!\varepsilon}^* \Biggl\{
\frac{g_{\rm L}^d\,\bar g_{\rm L}^{d'}}{4 c_{\rm w}^{}}
\Biggl[\Biggl(\frac{c_{\rm w}^2 k^2}{m_W^2}-1\Biggr)x_q^{}
- \frac{\bigl(16 m_W^2+k^2\bigr)c_{\rm w}^2 k^2}{3 m_W^4} \Biggr] P_{\rm L}^{}
\nonumber \\ && \hspace*{11ex} +\,\,
\frac{g_{\rm R}^d\,\bar g_{\rm R}^{d'}}{4 c_{\rm w}^{}}
\Biggl[ \Biggl(\frac{c_{\rm w}^2 k^2}{m_W^2}+7\Biggr)x_q^{}
- \frac{\bigl(16 m_W^2+k^2\bigr)c_{\rm w}^2 k^2}{3 m_W^4} \Biggr] P_{\rm R}^{}
\Biggr\} d ~,
\end{eqnarray}
where the $k^2$-dependent terms have been supplied by the second diagram in Fig.~\ref{penguins}.
Accordingly, setting  \,$k^2=m_Z^2=m_W^2/c_{\rm w}^2$,\,  one finds for \,$Z\to\bar d d'$\,
\begin{eqnarray} \label{z2dd}
{\cal M}_{Z\to\bar d d'}^{q,\rm div} &=&
\frac{g\,\Delta}{192\pi^2\,c_{\rm w}^3}\, \bar d'\!\!\not{\!\varepsilon} \Bigl[
-g_{\rm L}^d\,\bar g_{\rm L}^{d'}\,\bigl(16c_{\rm w}^2+1\bigr) P_{\rm L}^{}  \,+\,
g_{\rm R}^d\,\bar g_{\rm R}^{d'}\, \bigl(24c_{\rm w}^2\,x_q^{}-16c_{\rm w}^2-1\bigr) P_{\rm R}^{}
\Bigr] d ~.
\end{eqnarray}
Clearly, in the SM limit the $m_q^{}$-dependent part of the divergence in the \,$Z\to\bar d d'$\,
amplitude vanishes, but the remaining $x_q^{}$-independent part is removed only after applying
the GIM mechanism.
This is in accord with what is found in the literature~\cite{z2qq}.

In evaluating the box diagrams, we assume that the $W$ has SM couplings to leptons and
set the masses and momenta of the external fermions to zero.
Thus, from Fig.~\ref{boxes}(a) we obtain
\begin{eqnarray} \label{ddnn_box}
{\cal M}_{d\bar d'\to\nu\bar\nu}^{q\rm,box} &=&
\frac{g^2\, g_{\rm L}^d\,\bar g_{\rm L}^{d'}}{128\pi^2\,m_W^2} \Bigl[
\bigl(x_q^{}+x_\ell^{}-6\bigr) \Delta + {\cal B}_2^{}\bigl(x_q^{},x_\ell^{}\bigr) \Bigr]
\bar d'\gamma^\mu P_{\rm L}^{}d\, \bar\nu\gamma_\mu^{}P_{\rm L}^{}\nu
\nonumber \\ && \!\!\! +\,\,
\frac{g^2\, g_{\rm R}^d\,\bar g_{\rm R}^{d'}}{128\pi^2\,m_W^2} \Bigl[
\bigl(x_q^{}+x_\ell^{}-6\bigr) \Delta + {\cal B}_1^{}\bigl(x_q^{},x_\ell^{}\bigr) \Bigr]
\bar d'\gamma^\mu P_{\rm R}^{}d\, \bar\nu\gamma_\mu^{}P_{\rm L}^{}\nu  ~,
\end{eqnarray}
\begin{eqnarray} \label{ddll_box}
{\cal M}_{d\bar d'\to\ell^+\ell^-}^{q\rm,box} &=&
\frac{g^2\, g_{\rm L}^d\,\bar g_{\rm L}^{d'}}{128\pi^2\,m_W^2} \Bigl[
\bigl(6-x_q^{}\bigr) \Delta - {\cal B}_1^{}\bigl(x_q^{},0\bigr) \Bigr]
\bar d'\gamma^\mu P_{\rm L}^{}d\,\bar\ell\gamma_\mu^{} P_{\rm L}^{}\ell
\nonumber \\ && \!\! +\,\,
\frac{g^2\, g_{\rm R}^d\,\bar g_{\rm R}^{d'}}{128\pi^2\,m_W^2} \Bigl[
\bigl(6-x_q^{}\bigr) \Delta - {\cal B}_2^{}\bigl(x_q^{},0\bigr) \Bigr]
\bar d'\gamma^\mu P_{\rm R}^{}d\,\bar\ell\gamma_\mu^{} P_{\rm L}^{}\ell ~,
\end{eqnarray}
where
\begin{eqnarray} \label{b1b2}
\begin{array}{c}   \displaystyle
{\cal B}_1^{}(x,y) \,\,=\,\, \frac{3}{2}(x+y) + \frac{3(x+y-x y)}{(1-x)(1-y)}
+ \frac{\bigl(4 x^2-8 x^3+x^4\bigr)\, \ln x}{(y-x)(1-x)^2}
+ \frac{\bigl(4 y^2-8 y^3+y^4\bigr)\, \ln y}{(x-y)(1-y)^2} ~,
\vspace{2ex} \\   \displaystyle
{\cal B}_2^{}(x,y) \,\,=\,\, \frac{3}{2}(x+y) - \frac{9(x+y-x y)}{(1-x)(1-y)} +
\frac{(4-x)^2 x^2\, \ln x}{(y-x)(1-x)^2} + \frac{(4-y)^2 y^2\, \ln y}{(x-y)(1-y)^2} ~,
\end{array}
\end{eqnarray}
and  $\ell$ in Eq.~(\ref{ddnn_box}) refers to the charged lepton in the loop.
Since the divergences in these box contributions depend on $m_q^{}$, they cannot be removed
by the GIM mechanism in the SM limit, Eq.~(\ref{smlimit}).
This is different from the corresponding contributions computed in the $R_\xi$
gauge, which are finite~\cite{Inami:1980fz}

Combining the \,$d\bar d'\to Z^*\to\nu\bar\nu$\, amplitude derived from Eq.~(\ref{ddz}) and
the box-diagram amplitude in Eq.~(\ref{ddnn_box}), we then find
\begin{eqnarray} \label{Mdd2nn}
{\cal M}_{d\bar d'\to\nu\bar\nu}^q &=&
\frac{g^2\, g_{\rm L}^d\,\bar g_{\rm L}^{d'}}{32\pi^2\,m_W^2} \Biggl[
\frac{\bigl(x_\ell^{}-6\bigr)\Delta}{4} + 4 X_0^{}\bigl(x_q^{}\bigr)
+ \frac{{\cal B}_2^{}\bigl(x_q^{},x_\ell^{}\bigr)-{\cal B}_2^{}\bigl(x_q^{},0\bigr)}{4}
\Biggr] \bar d'\gamma^\mu P_{\rm L}^{}d\, \bar\nu\gamma_\mu^{}P_{\rm L}^{}\nu
\nonumber \\ && \hspace*{-5ex} +\,\,
\frac{g^2\, g_{\rm R}^d\,\bar g_{\rm R}^{d'}}{32\pi^2\,m_W^2} \Biggl[
\biggl(2 x_q^{}+\frac{x_\ell^{}-6}{4}\biggr)\Delta + \tilde X\bigl(x_q^{}\bigr)
+ \frac{{\cal B}_1^{}\bigl(x_q^{},x_\ell^{}\bigr)-{\cal B}_1^{}\bigl(x_q^{},0\bigr)}{4}
\Biggr] \bar d'\gamma^\mu P_{\rm R}^{}d\, \bar\nu\gamma_\mu^{}P_{\rm L}^{}\nu ~,
\hspace*{4ex}
\end{eqnarray}
where
\begin{eqnarray} \label{x0}
X_0^{}(x) \,\,=\,\, \frac{x(x+2)}{8(x-1)} + \frac{3x(x-2)}{8(x-1)^2}\,\ln x ~, \hspace{7ex}
\tilde X(x) \,\,=\,\, 2 x - \frac{5 x-2 x^2}{1-x}\,\ln x - 4 X_0^{}(x) ~.
\end{eqnarray}
In writing down this amplitude, we have neglected the contribution of the $Z$-magnetic dipole
terms, with $Z_{\rm L,R}^q$, in Eq.~(\ref{ddz}), which are small compared to the $z_{\rm L,R}^q$
terms due to the $m_W^{}$ suppression.
We can see that in the SM limit, Eq.~(\ref{smlimit}), the divergent part of
${\cal M}_{d\bar d'\to\nu\bar\nu}^q$ no longer depends on~$m_q^{}$ because of cancelation
between the $m_q^{}$-dependent divergent terms in the $Z$-penguin and box contributions.
The remaining divergence will then disappear when the GIM mechanism operates.

For  \,$d\bar d'\to\ell^+\ell^-$,\,  the amplitude receives not only the $Z$-mediated
and box contributions, but also the photon-mediated one derived from Eq.~(\ref{ddf}).
As a result
\begin{eqnarray} \label{Mdd2ll}
{\cal M}_{d\bar d'\to\ell^+\ell^-}^q &=&
\frac{g^2\, g_{\rm L}^d\,\bar g_{\rm L}^{d'}}{32\pi^2\,m_W^2} \biggl(
\frac{3\Delta}{2} \,-\, 4 Y_0^{}\bigl(x_q^{}\bigr)\biggr)
\bar d'\gamma^\mu P_{\rm L}^{}d\, \bar\ell\gamma_\mu^{}P_{\rm L}^{}\ell
\nonumber \\ && \!\! +\,\,
\frac{g^2\, g_{\rm L}^d\,\bar g_{\rm L}^{d'}\,s_{\rm w}^2}{32\pi^2\,m_W^2} \biggl(
-\frac{8\Delta}{3} \,+\, 8 Z_0^{}\bigl(x_q^{}\bigr)\biggr)
\bar d'\gamma^\mu P_{\rm L}^{}d\, \bar\ell\gamma_\mu^{}\ell
\nonumber \\ && \!\! +\,\,
\frac{g^2\, g_{\rm R}^d\,\bar g_{\rm R}^{d'}}{32\pi^2\,m_W^2} \biggl[
\biggl(\frac{3}{2}-2 x_q^{}\biggr)\Delta \,+\, \tilde Y\bigl(x_q^{}\bigr) \biggr]
\bar d'\gamma^\mu P_{\rm R}^{}d\, \bar\ell\gamma_\mu^{}P_{\rm L}^{}\ell
\nonumber \\ && \!\! +\,\,
\frac{g^2\, g_{\rm R}^d\,\bar g_{\rm R}^{d'}\,s_{\rm w}^2}{32\pi^2\,m_W^2} \biggl[
\biggl(4 x_q^{}-\frac{8}{3}\biggr)\Delta \,+\, \tilde Z\bigl(x_q^{}\bigr) \biggr]
\bar d'\gamma^\mu P_{\rm R}^{}d\, \bar\ell\gamma_\mu^{}\ell
\nonumber \\ && \!\! +\,\,
\frac{i g^2\,s_{\rm w}^2}{16\pi^2\,m_W^2\, k^2}\, \bar d' \sigma^{\mu\nu}k_\nu^{}
\bigl(F_{\rm L}^q P_{\rm L}^{}+F_{\rm R}^q P_{\rm R}^{}\bigr)d\, \bar\ell\gamma_\mu^{}\ell ~,
\end{eqnarray}
where
\begin{eqnarray} \label{y0z0}
\begin{array}{c}   \displaystyle
Y_0^{}(x) \,\,=\,\, \frac{x(x-4)}{8(x-1)} + \frac{3x^2}{8(x-1)^2}\,\ln x ~,
\vspace{2ex} \\   \displaystyle
Z_0^{}(x) \,\,=\,\, \frac{18x^4-163x^3+259x^2-108x}{144(x-1)^3} +
\frac{24x^4-6x^3-63x^2+50x-8}{72(x-1)^2}\,\ln x ~,
\end{array}
\end{eqnarray}
\begin{eqnarray}
\begin{array}{c}   \displaystyle
\tilde Y(x) \,\,=\,\, - 2 x + \frac{5 x-2 x^2}{1-x}\,\ln x + 4 Y_0^{}(x) ~, \hspace{7ex}
\tilde Z(x) \,\,=\,\, 2x - 4 x\,\ln x + 8 Z_0^{}(x) ~,
\end{array}
\end{eqnarray}
and we have again neglected the contribution of the $Z$-magnetic dipole terms in Eq.~(\ref{ddz}).
In the SM limit, the divergent terms in the total amplitude are also independent of $m_q^{}$
due to cancelation among the $m_q^{}$-dependent divergent parts of the penguin and box
contributions.
This, along with the similar cancelation in the \,$d\bar d'\to\nu\bar\nu$\, case, explicitly
confirms the result of unitary-gauge analysis in Ref.~\cite{Kuksa:2003ga}.

In terms of the couplings $\kappa_{qd}^{\rm L,R}$ parameterizing the anomalous
interactions of quarks and the $W$ boson considered in Ref.~\cite{He:2009hz}, we have
\begin{eqnarray} \label{gLgR}
g_{\rm L}^d \,\,=\,\, \frac{g}{\sqrt2}\, V_{qd}^{}\bigl(1+\kappa_{qd}^{\rm L}\bigr) ~, \hspace{7ex}
g_{\rm R}^d \,\,=\,\, \frac{g}{\sqrt2}\, V_{qd}^{}\kappa_{qd}^{\rm R} ~,
\end{eqnarray}
where the $\kappa$'s are assumed to be small compared to unity.
It is then straightforward to arrive at the effective Hamiltonians for
\,$d\bar d'\to\nu\bar\nu$\, and \,$d\bar d'\to\ell^+\ell^-$\, within the SM,
as well as those induced by the anomalous couplings, given in Ref.~\cite{He:2009hz}.

For \,$d\bar d'\to\bar d d'$,\, we derive the amplitude from the two box-diagrams in
Fig.~\ref{boxes} with quarks $d$ and \,$d'\neq d$\,  in the external legs and quarks $q$
and~$q'$ in the loops.  In this case we adopt the parametrization in Eq.~(\ref{gLgR}).
With the contribution of the anomalous couplings included to second order in $\kappa$,
it follows that
\begin{eqnarray} \label{dd'2dd'}
{\cal M}_{d\bar d'\to\bar d d'}^q &=&
\frac{g^4\, \lambda_q^{\rm L}\lambda_{q'}^{\rm L}}{256\pi^2\,m_W^2} \Biggl[
\bigl(6-x_q^{}-x_{q'}^{}\bigr)\Delta \,-\, {\cal B}_1^{}\bigl(x_q^{},x_{q'}^{}\bigr)
\Biggr] \bar d_1'\gamma^\alpha P_{\rm L}^{}d_1^{}\,\bar d_2'\gamma_\alpha^{}P_{\rm L}^{}d_2^{}
\nonumber \\ && \!\! +\,\,
\frac{g^4\, \lambda_q^{}\lambda_{q'}^{}}{256\pi^2\,m_W^2}\Biggl[
\bigl(6-x_q^{}-x_{q'}^{}\bigr)\Delta \,-\, {\cal B}_2^{}\bigl(x_q^{},x_{q'}^{}\bigr) \Biggr]
\nonumber \\ && \,\,\times\,\, \Bigl( \kappa_{qd'}^{\rm R*}\kappa_{qd}^{\rm R}\,
\bar d_1'\gamma^\alpha P_{\rm L}^{}d_1^{}\,\bar d_2'\gamma_\alpha^{}P_{\rm R}^{}d_2^{}
+ \kappa_{q'd'}^{\rm R*}\kappa_{q'd}^{\rm R}\,
\bar d_1'\gamma^\alpha P_{\rm R}^{}d_1^{}\,\bar d_2'\gamma_\alpha^{}P_{\rm L}^{}d_2^{} \Bigr)
\nonumber \\ && \!\! -\,\,
\frac{g^4\,\lambda_q^{}\lambda_{q'}^{}\,m_q^{}m_{q'}^{}}{64\pi^2\,m_W^4} \Bigl(
\Delta + {\cal B}_3^{}\bigl(x_q^{},x_{q'}^{}\bigr) \Bigr) \Bigl(
\kappa_{qd}^{\rm R}\kappa_{q'd}^{\rm R}\, \bar d_1'P_{\rm R}^{}d_1^{}\,\bar d_2'P_{\rm R}^{}d_2^{}
+ \kappa_{qd'}^{\rm R*}\kappa_{q'd'}^{\rm R*}\, \bar d_1'P_{\rm L}^{}d_1^{}\,
\bar d_2'P_{\rm L}^{}d_2^{} \Bigr)
\nonumber \\ && \!\! +\,\, \bigl(d_1'\leftrightarrow d_2'\bigr) ~,
\end{eqnarray}
where we have distinguished the two $d^{(\prime)}$'s,
\begin{eqnarray}
\lambda_q^{\rm L} \,\,=\,\, \lambda_q^{}\,
\bigl(1+\kappa_{qd'}^{\rm L*}\bigr)\bigl(1+\kappa_{qd}^{\rm L}\bigr) ~, \hspace{7ex}
\lambda_q^{\rm R}\,\,=\,\,\lambda_q^{}\,\kappa_{qd'}^{\rm R*}\kappa_{qd}^{\rm R} ~,
\end{eqnarray}
\begin{eqnarray} \label{b3}
{\cal B}_3^{}(x,y) \,\,=\,\,
\frac{x y-x-y-2}{(1-x)(1-y)} + \frac{\bigl(4 x-2 x^2+x^3)\,\ln x}{(y-x)(1-x)^2} +
\frac{\bigl(4 y-2 y^2+y^3\bigr)\,\ln y}{(x-y)(1-y)^2} ~,
\end{eqnarray}
and we have neglected finite terms quadratic in $\kappa^{\rm R}$ containing
\,$\bar d'\sigma^{\mu\nu}P_{\rm L,R}^{}d\,\bar d'\sigma_{\mu\nu}^{}P_{\rm L,R}^{}d$\,  whose
contributions to the neutral-meson mixing would vanish in the vacuum saturation approximation.
It is also straightforward to obtain the corresponding effective Hamiltonians given in
Ref.~\cite{He:2009hz}.
This \,$d\bar d'\to\bar d d'$\, amplitude in unitary gauge has been previously calculated in
Ref.~\cite{Lee:2008xr}, but our result for the $\kappa^{\rm R}$ terms disagrees with theirs.

Finally, we remark that some of the physical amplitudes we have found above generated by
the anomalous couplings contain divergent parts.
As discussed in Ref.~\cite{He:2009hz}, this is due to the fact that the effective theory
with anomalous couplings in Eq.~(\ref{Ludw}) is not renormalizable, and the divergences are
understood in the context of effective field theories as contributions to the coefficients of
higher-dimension operators, which are not included in our analysis.

In conclusion, we have provided the detailed derivation of loop formulas summarized in
our recent work evaluating several two- and four-fermion FCNC transitions induced at one-loop
level by the anomalous charm-$W$ couplings.
In treating the loop diagrams, we have worked in unitary gauge and used dimensional
regularization to handle the divergences arising in some of the loops.
We have compared our results with the corresponding ones obtained in the literature
using unitary and $R_\xi$ gauges, where available.
In particular, we have discussed subtleties in the cancelation of divergences from individual
contributions to some of the physical amplitudes and derived expressions consistent with
those obtained in $R_\xi$ gauges.
Finally, we have provided, perhaps for the first time, the unitary-gauge expressions for
the separate diagrams contributing to the \,$d\to d'{\cal V}^*$\, amplitude.
These particular results can be applied to other models involving fermions and gauge bosons
with interactions similar in form to those considered here.
For example, the contribution of the $Z'$ boson to the \,$b\to s\gamma$\, transition can be
easily found from the magnetic terms in our general expression for the \,$d\to d'{\cal V}^*$\,
amplitude, with appropriate couplings.

\acknowledgments
The work of X.G.H. and J.T. was supported in part by NSC and NCTS.
The work of G.V. was supported in part by DOE under contract number DE-FG02-01ER41155.

\appendix

\section{Amplitudes for \boldmath$d\to d'\cal V^*$\label{ddV}}

The Lagrangian describing a neutral gauge-boson $\cal V$ coupling to a fermion-antifermion
pair, $f\bar f$, or a~$W$-boson pair can be expressed as
\begin{eqnarray} \label{Lv}
{\cal L}_{\cal V}^{} \,\,=\,\, -\bar f\gamma^\mu\bigl(L_f^{\cal V}P_{\rm L}^{}
+ R_f^{\cal V}P_{\rm R}^{}\bigr)f\,{\cal V}_\mu^{} \,+\,
i g_W^{\cal V} \bigl[ {\cal V}^\mu \bigl( W^+_{\mu\nu} W^{-\nu} - W^{+\nu} W^-_{\mu\nu} \bigr)
- {\cal V}^{\mu\nu} W^+_\mu  W^-_\nu \bigr] ~,
\end{eqnarray}
where  $L_f^{\cal V}$, $R_f^{\cal V}$, and $g_W^{\cal V}$ contain the coupling constants,
and  \,$X_{\mu\nu}=\partial_\mu X_\nu-\partial_\nu X_\mu$.\,
This and ${\cal L}_{UDW}$ in Eq.~(\ref{Ludw}) provide the relevant vertices for the
$d\to d'\cal V^*$ transition at one-loop level, leading to the diagrams in Fig.~\ref{penguins},
where the external quarks $d$ and $d'$ are on-shell and $\cal V$ is off-shell.
The resulting amplitudes for the contributions (a,b,c), with $q$ being the quark in the loops
and $\cal V$ having a four-momentum~$k$, are, respectively,
\begin{eqnarray} \label{ddv_a}
{\cal M}_{d\to d'{\cal V}^*}^{{\rm(a)}q} &=&
\frac{g_{\rm L}^d\, \bar g_{\rm L}^{d'}}{16\pi^2}\, \bar d' \Biggl[
\Biggl( \frac{R_q^{\cal V}-4L_q^{\cal V}}{2}\, x_q^{}\,\Delta \,+\,
L_q^{\cal V}\, V_1^{}\bigl(x_q^{}\bigr) + R_q^{\cal V}\, V_2^{}\bigl(x_q^{}\bigr) \Biggr)
\!\!\not{\!\varepsilon}^* P_{\rm L}^{}
\nonumber \\ && \hspace*{9ex} +\,\,
\Bigl(L_q^{\cal V}\, V_3^{}\bigl(x_q^{}\bigr)+R_q^{\cal V}\, V_4^{}\bigl(x_q^{}\bigr) \Bigr)
\frac{k^2 \!\not{\!\varepsilon}^*-\!\!\not{\!k}\,\varepsilon^*\!\cdot\!k}{m_W^2}\,P_{\rm L}^{}
\nonumber \\ && \hspace*{9ex} +\,\,
\Bigl(L_q^{\cal V}\,V_5^{}\bigl(x_q^{}\bigr)+R_q^{\cal V}\,V_6^{}\bigl(x_q^{}\bigr)\Bigr)
\bigl(m_{d'}^{}\,P_{\rm L}^{}+m_d^{}\,P_{\rm R}^{}\bigr)
\frac{i\sigma^{\mu\nu}\varepsilon_\mu^*k_\nu^{}}{m_W^2}
\nonumber \\ && \hspace*{9ex} +\,\,
\frac{\bar g_{\rm R}^{d'}}{\bar g_{\rm L}^{d'}} \bigl(L_q^{\cal V}+R_q^{\cal V}\bigr)\,
V_7^{}\bigl(x_q^{}\bigr)\, m_q^{}\,P_{\rm L}^{}\,
\frac{i\sigma^{\mu\nu}\varepsilon_\mu^*k_\nu^{}}{m_W^2}
\Biggr] d
\nonumber \\ && \! +\,\,
\bigl( g_{\rm L}^{}\leftrightarrow g_{\rm R}^{},\, L_q^{\cal V}\leftrightarrow R_q^{\cal V},\,
P_{\rm L}^{}\leftrightarrow P_{\rm R}^{} \bigr) ~,
\end{eqnarray}
\begin{eqnarray} \label{ddv_b}
{\cal M}_{d\to d'{\cal V}^*}^{{\rm(b)}q} &=&
\frac{g_W^{\cal V}}{16\pi^2}\, \bar d' \Biggl\{ \biggl( \frac{3 x_q^{}\,\Delta}{2} \,-\,
V_1^{}\bigl(x_q^{}\bigr)-V_2^{}\bigl(x_q^{}\bigr) \biggr) \!\not{\!\varepsilon}^*
\bigl( g_{\rm L}^d\, \bar g_{\rm L}^{d'} P_{\rm L}^{} +
       g_{\rm R}^d\, \bar g_{\rm R}^{d'} P_{\rm R}^{} \bigr)
\nonumber \\ && \hspace*{9ex} +\,\,
\biggl( \frac{3 x_q^{}-16}{12}\,\Delta \,+\, V_8^{}\bigl(x_q^{}\bigr) \biggr)
\frac{k^2\!\not{\!\varepsilon}^*-\!\!\not{\!k}\, \varepsilon^*\!\cdot\!k}{m_W^2}
\bigl( g_{\rm L}^d\, \bar g_{\rm L}^{d'} P_{\rm L}^{} +
       g_{\rm R}^d\, \bar g_{\rm R}^{d'} P_{\rm R}^{} \bigr)
\nonumber \\ && \hspace*{9ex} +\,\,
V_9^{}\bigl(x_q^{}\bigr)\, \bigl[ \bigl(g_{\rm L}^d\,\bar g_{\rm L}^{d'}m_{d'}^{}
+ g_{\rm R}^d\,\bar g_{\rm R}^{d'}m_d^{}\bigr) P_{\rm L}^{}
+ \bigl( g_{\rm L}^d\, \bar g_{\rm L}^{d'} m_d^{}
+ g_{\rm R}^d\, \bar g_{\rm R}^{d'} m_{d'}^{}\bigr) P_{\rm R}^{} \bigr]
\frac{i\sigma^{\mu\nu}\varepsilon_\mu^* k_\nu^{}}{m_W^2}
\nonumber \\ && \hspace*{9ex} +\,\,
V_{10}^{}\bigl(x_q^{}\bigr)\, m_q^{}
\bigl(g_{\rm L}^d\, \bar g_{\rm R}^{d'} P_{\rm L}^{} +
g_{\rm L}^{d'}\, \bar g_{\rm R}^d P_{\rm R}^{}\bigr)
\frac{i\sigma^{\mu\nu} \varepsilon_\mu^*k_\nu^{}}{m_W^2} \Biggr\} d ~,
\end{eqnarray}
\begin{eqnarray} \label{ddv_c}
{\cal M}_{d\to d'{\cal V}^*}^{{\rm(c)}q} &=& \frac{1}{16\pi^2} \biggl(\frac{3 x_q^{}\,\Delta}{2}
\,-\, V_1^{}\bigl(x_q^{}\bigr)-V_2^{}\bigl(x_q^{}\bigr) \biggr) \bar d'\!\!\not{\!\varepsilon}^*
\bigl( g_{\rm L}^d\, \bar g_{\rm L}^{d'} L_d^{\cal V} P_{\rm L}^{} +
       g_{\rm R}^d\, \bar g_{\rm R}^{d'} R_d^{\cal V} P_{\rm R}^{} \bigr) d ~,
\end{eqnarray}
where  $\Delta$ is given in Eq.~(\ref{delta}),
$\varepsilon$ and $k$ are the polarization and four-momentum of $\cal V$, respectively,
\begin{eqnarray}
\bar g_{\rm L}^{d'} \,\,=\,\, \bigl(g_{\rm L}^{d'}\bigr)^* ~, \hspace{5ex}
\bar g_{\rm R}^{d'} \,\,=\,\, \bigl(g_{\rm R}^{d'}\bigr)^* ~, \hspace{5ex}
x_f^{} \,\,=\,\, \frac{m_f^2}{m_W^2} ~, \hspace{5ex}
\sigma^{\mu\nu} \,\,=\,\, \frac{i}{2}[\gamma^\mu,\gamma^\nu] ~,
\end{eqnarray}
\begin{eqnarray}
\begin{array}{c}   \displaystyle
V_1^{}(x) \,\,=\,\, -\frac{3}{2} - x + 2x\,\ln x ~, \hspace{5ex}
V_2^{}(x) \,\,=\,\, \frac{-7 x+x^2}{4(1-x)} - \frac{4 x-2 x^2+x^3}{2(1-x)^2}\,\ln x ~,
\vspace{2ex} \\   \displaystyle
V_3^{}(x) \,\,=\,\, -\frac{2}{9} + \frac{10 x-5 x^2-11 x^3}{36(1-x)^3} -
\frac{(2-3 x)^2}{6(1-x)^4}\,\ln x ~,
\vspace{2ex} \\   \displaystyle
V_4^{}(x) \,\,=\,\, \frac{1}{3} + \frac{11 x-7 x^2+2 x^3}{9(1-x)^3} +
\frac{2 x\,\ln x}{3(1-x)^4} ~,
\vspace{2ex} \\   \displaystyle
V_5^{}(x) \,\,=\,\, \frac{2}{3} + \frac{5 x-22 x^2-5 x^3}{12(1-x)^3} +
\frac{x-3 x^2}{2(1-x)^4}\,\ln x ~, \hspace{5ex}
V_6^{}(x) \,\,=\,\, \frac{1}{4} -  \frac{3}{4}\, V_4^{}(x)
\vspace{2ex} \\   \displaystyle
V_7^{}(x) \,\,=\,\, -1 - \frac{9 x-3 x^2}{4(1-x)^2} - \frac{3 x\,\ln x}{2(1-x)^3} ~,
\vspace{2ex} \\   \displaystyle
V_8^{}(x) \,\,=\,\, \frac{25}{18} + \frac{27 x-113 x^2+83 x^3-9 x^4}{24(1-x)^3} -
\frac{31 x^3-28 x^4+3 x^5}{12(1-x)^4}\,\ln x ~,
\vspace{2ex} \\   \displaystyle
V_9^{}(x) \,\,=\,\, \frac{5}{6} - \frac{x-5 x^2-2 x^3}{4(1-x)^3} +
\frac{3 x^3\,\ln x}{2(1-x)^4} ~, \hspace{5ex}
V_{10}^{}(x) \,\,=\,\, 6 V_4^{}(x)+V_5^{}(x)+3 V_7^{}(x)-2 V_9^{}(x) ~.
\end{array}
\end{eqnarray}
In obtaining these results, we have assumed that $k$ and $m_{d,d'}$ are small compared
to~$m_W^{}$, kept terms to second order in $k$, and taken at the end of the calculation
the usual limit of vanishing  $m_{d,d'}$ for the nonleading terms~\cite{Ma:1979px,Inami:1980fz}.
It is worth noting that the $x_q^{}$-dependent part of the divergent term in each of
these amplitudes arises from the $p^\mu p^\nu$ terms of the $W$ propagators.

The separate expressions above for ${\cal M}_{d\to d'{\cal V}^*}^{{\rm(a,b,c)}q}$ in unitary
gauge may have been written down for the first time in this work.
They are general and follow from any effective interactions of the form given in
Eqs.~(\ref{Ludw}) and~(\ref{Lv}), provided that the masses of the external fermions and
the momentum of the external gauge boson are much less than the gauge-boson mass in the loops.
In the case of the standard model, the couplings of \,${\cal V}=Z,\gamma,g_a^{}$\, in
${\cal L}_{\cal V}^{}$ are parametrized by
\begin{eqnarray} \label{sm}
\begin{array}{c}   \displaystyle
L_f^Z \,\,=\,\, \frac{g}{c_{\rm w}^{}}\bigl(I_{3f}^{}-s_{\rm w}^2\,Q_f^{}\bigr) ~, \hspace{5ex}
R_f^Z \,\,=\,\, -\frac{g}{c_{\rm w}^{}}\, s_{\rm w}^2\, Q_f^{} ~,
\vspace{2ex} \\   \displaystyle
L_f^\gamma \,\,=\,\, R_f^\gamma \,\,=\,\, e Q_f^{} ~, \hspace{5ex}
L_f^{g_a^{}} \,\,=\,\, R_f^{g_a^{}} \,\,=\,\, g_{\rm s}^{} t_a^{} ~,
\vspace{2ex} \\   \displaystyle
g_W^Z \,\,=\,\, c_{\rm w}^{}\, g ~, \hspace{5ex}
g_W^\gamma \,\,=\,\, e \,\,=\,\, g\, s_{\rm w}^{} ~, \hspace{5ex}
g_W^{g_a^{}} \,\,=\,\, 0 ~,
\vspace{2ex} \\   \displaystyle
c_{\rm w}^{} \,\,=\,\, \cos\theta_{\rm W}^{} ~, \hspace{5ex}
s_{\rm w}^{} \,\,=\,\, \sin\theta_{\rm W}^{} ~,
\end{array}
\end{eqnarray}
where $I_{3f}$ and $Q_f$ are as usual the third component of weak isospin of $f$ and its
electric charge, respectively, and $t_a^{}$ are the color-SU(3) generators satisfying
\,${\rm Tr}(t_a t_b)=\frac{1}{2}\delta_{ab}$.\,

\end{document}